\input harvmac
\def\ra{\rightarrow}
\def\O{{\cal O}}
\def\gev{\rm GeV}
\def\tev{\rm TeV}
\def\mpl{m_{\rm Pl}}
\def\ea{\epsilon_1}
\def\eb{\epsilon_2}
\def\ec{\epsilon_3}
\def\ef{\epsilon_3}
\def\H{{\cal H}}
\def\gsim{{~\raise.15em\hbox{$>$}\kern-.85em
          \lower.35em\hbox{$\sim$}~}}
\def\lsim{{~\raise.15em\hbox{$<$}\kern-.85em
          \lower.35em\hbox{$\sim$}~}}
\Title{hep-ph/9603233, RU-96-11, WIS-96/11/Feb-PH}
{\vbox{\centerline{Solving the Supersymmetric CP Problem}
 \centerline{with Abelian Horizontal Symmetries}}}
\bigskip
\centerline{Yosef Nir}
\smallskip
\centerline{\it Department of Particle Physics}
\centerline{\it Weizmann Institute of Science, Rehovot 76100, Israel}
\bigskip
\centerline{Riccardo Rattazzi}
\smallskip
\centerline{\it Department of Physics and Astronomy}
\centerline{\it Rutgers University, Piscataway, NJ 08855}
\bigskip
\baselineskip 18pt

\noindent
Models that combine Abelian horizontal symmetries and
spontaneous CP violation can (i) explain the smallness
and hierarchy in quark parameters; (ii) satisfactorily
suppress supersymmetric contributions to flavor changing
neutral current processes; (iii) solve the $\mu$-problem;
and (iv) suppress supersymmetric contributions to CP
violating observables to an acceptable level.
The CKM phase is  $\O(1)$ and responsible, through
Standard Model box diagrams, to $\epsilon_K$.
The supersymmetric CP violating phases are suppressed,
$\phi_A\sim\lambda^6$ and $\phi_B\sim\lambda^{8}$
($\lambda\sim0.2$), leading to an electric dipole moment
of the neutron that is about 2--3 orders of magnitude
below the experimental bound.

\Date{2/96}

\newsec{Introduction}
Supersymmetric theories introduce new sources of CP violation.
Even with just the minimal supersymmetric extension of the
Standard Model, there are two new phases
\ref\DGH{M. Dugan, B. Grinstein and L. Hall,
 Nucl. Phys. B255 (1985) 413.}
\ref\DiTh{S. Dimopoulos and S. Thomas, hep-ph/9510220.}:
\eqn\phiAB{\eqalign{
\phi_A=&\ \arg\left(A^* m_\lambda\right),\cr
\phi_B=&\ \arg\left(m_\lambda\mu m_{12}^{2*}\right),\cr}}
where $A$ and $m_{12}^2$ are the coefficients of,
respectively, the trilinear
and bilinear soft SUSY breaking terms and $m_\lambda$
is the gaugino mass. Unless these phases are
$\lsim\O(10^{-2})$, or supersymmetric masses are
$\gsim\O(1\ \tev)$, the supersymmetric contribution to
the electric dipole moment of the neutron $d_N$ is well
above the experimental bound. This is the Supersymmetric
CP Problem. Furthermore, the CKM phase $\delta_{KM}$ contributes
to $K-\bar K$ mixing through many new diagrams
involving supersymmetric particles. For generic
squark masses, these contributions are well above
the experimental value of $\epsilon_K$.

Horizontal symmetries, invoked to explain the smallness
and hierarchy in fermion masses and mixing angles,
have further interesting implications in the
supersymmetric framework. In particular, they
constrain the form of the mass-squared matrices
of squarks and sleptons and, consequently, are
able to solve or, at least, relax the problems of
supersymmetric flavor changing neutral currents
(FCNC). This idea has been investigated for both Abelian
\nref\NirSeiberg{Y. Nir and N. Seiberg, Phys. Lett.
 B309 (1993) 337.}%
\nref\Sequel{M. Leurer, Y. Nir and N. Seiberg,
 Nucl. Phys. B420 (1994) 468.}%
\refs{\NirSeiberg-\Sequel}\ and non-Abelian
\nref\DKL{M. Dine, A. Kagan and R.G. Leigh,
 Phys. Rev. D48 (1993) 4269.}%
\nref\PoTo{A. Pomarol and D. Tommasini, hep-ph/9507462.}%
\nref\HaMu{L.J. Hall and H. Murayama, Phys. Rev. Lett. 75 (1995)
3985.}%
\nref\BDH{R. Barbieri, G. Dvali and L. Hall, hep-ph/9512388.}%
\nref\CHM{C.D. Carone, L.J. Hall and H. Murayama,
 hep-ph/9512399.}%
\refs{\DKL-\CHM}\ symmetries.
In both frameworks, even the most stringent FCNC constraints--
$\Delta m_K$ in the quark sector and $\mu\ra e\gamma$ in the
lepton sector --can be satisfied: an Abelian symmetry (in
combination with holomorphy) can precisely align the fermion
mass matrix with the corresponding sfermion mass-squared
matrix (leading to highly suppressed gaugino mixing angles),
while a non-Abelian symmetry can lead to degeneracy between
the first two sfermion generations.

In this work, we investigate whether the Abelian symmetries
that lead to satisfactory quark-squark alignment may
simultaneously solve the SUSY CP problems.\foot{For a related
study, in the framework of non-Abelian symmetries, see \PoTo.}
Indeed, it has already been shown in \refs{\NirSeiberg-\Sequel}\ that the
quark-squark alignment could be precise enough so that
the magnitude of the SUSY contribution to $K-\bar K$ mixing
is orders of magnitude below the Standard Model box diagrams.
The contribution to $\epsilon_K$ is then very small, even
for $\O(1)$ phases . However, the contributions to the
electric dipole moment of the neutron from phases of the
type $\phi_A,\phi_B$ are, in general, not suppressed below
those of generic Supersymmetric models. An extra ingredient,
beyond the horizontal symmetry, is required. We here show
that the required suppression can be achieved when CP breaking
is spontaneous. Our basic assumption is that CP is preserved by the 
sector responsible for supersymmetry breaking, while it is spontaneously
broken in the flavor sector.
 (For studies of spontaneous CP violation in
various supersymmetric models, see refs.
\nref\Maek{N. Maekawa, Phys. Lett. B282 (1992) 387.}%
\nref\Poma{A. Pomarol, Phys. Lett. B287 (1992) 331.}%
\nref\Pomb{A. Pomarol, Phys. Rev. D47 (1993) 273.}%
\nref\BaBa{K.S. Babu and S.M. Barr, Phys. Rev. D49 (1994) 2156.}%
\nref\MaRa{M. Masip and A. Rasin, Phys. Rev. D52 (1995) 3768.}%
\nref\MaRb{M. Masip and A. Rasin, hep-ph/9508365.}%
\nref\BaBb{K.S. Babu and S.M. Barr, Phys. Rev. Lett. 72 (1994) 2831.}%
\refs{\Maek-\BaBb}.)

Below, we present an explicit model. This is a minimal
extension of the quark-squark alignment models of ref. \Sequel\
that can accommodate spontaneous CP violation. In our
conclusions, we point out which of the ingredients
in this model might apply in a more general framework.

\newsec{The Model}
The model of ref. \Sequel\ assumed an Abelian horizontal symmetry
\eqn\Symmetry{\H=U(1)_1\times U(1)_2.}
The symmetry is spontaneously broken by the VEVs of two
Standard Model gauge singlets, $S_1$ and $S_2$, with $\H$-charges
\eqn\Scharges{S_1(-1,0),\ \ \ S_2(0,-1).}
The information about the horizontal symmetry breaking
is communicated to the observed quarks at a high energy
scale, possibly the Planck scale $m_{\rm Pl}$, thus
providing two small breaking parameters (this is the
Froggatt-Nielsen mechanism
\nref\FrNi{C.D. Froggatt and H.B. Nielsen,
 Nucl. Phys. B147 (1979) 277.}%
\nref\LeNiSe{M. Leurer, Y. Nir and N. Seiberg, Nucl. Phys. B398
(1993) 319.}%
\refs{\FrNi-\LeNiSe}):
\eqn\breaking{\epsilon_1\equiv{\vev{S_1}\over m_{\rm Pl}}\sim
\lambda,\ \ \ \epsilon_2\equiv{\vev{S_2}\over m_{\rm Pl}}\sim
\lambda^2,}
where $\lambda$ is taken to be of the order of the Cabibbo
angle, $\lambda\sim0.2$. With this scalar content, it is
impossible to have spontaneous CP violation, because any
phase in $\vev{S_i}$ can be rotated away by means of a
$U(1)_i$ rotation.
In order to have spontaneous CP violation, at least
one additional singlet $S_3$ is required, which transforms
under either or both of $U(1)_1$ and $U(1)_2$. We choose
\eqn\Sthree{S_3(-3,-1),\ \ \ \epsilon_3\equiv{\vev{S_3}\over
m_{\rm Pl}}\sim\lambda^5.}
Without loss of generality, we can take $\epsilon_1$ and
$\epsilon_2$ real, while $\epsilon_3$ is complex.
(Both $|\vev{S_3}/(\vev{S_1}^3\vev{S_2})|\sim1$ and
$\arg[\vev{S_3}/(\vev{S_1}^3\vev{S_2})]\sim1$
will be shown to arise naturally.)

We assign the following $\H$ charges to the matter supermultiplets
($Q_i$ are quark doublets, $\bar d_i$ and $\bar u_i$ are
quark singlets, $\phi_u$ and $\phi_d$ are the Higgs doublets):
\eqn\mattercharges{\matrix{
Q_1&Q_2&Q_3&&\bar d_1&\bar d_2&\bar d_3\cr
(3,0)&(0,1)&(0,0)&&(-2,3)&(5,-1)&(1,1)\cr
\bar u_1&\bar u_2&\bar u_3&&&\phi_u&\phi_d\cr
(-1,2)&(1,0)&(0,0)&&&(0,0)&(-1,0)\cr}}

To find the quark mass matrices and the squark mass-squared
matrices, we note that the following selection rules hold
in the effective theory below the Planck scale:
\item{(i)} Terms in the superpotential that carry charge
$(n_1,n_2)$ are suppressed by $\lambda^{n_1+2n_2}$ if
$n_1\geq0$ and $n_2\geq0$ and vanish otherwise.
\item{(ii)} Terms in the Kahler potential that carry charge
$(n_1,n_2)$ are suppressed by $\lambda^{|n_1|+2|n_2|}$.

We now present the form of the various mass matrices
for quarks and squarks that follow
from the selection rules in our specific model.  We emphasize
that in each entry in the mass matrices below, we omit
an unknown coefficient of order 1. However, assuming that
the only source of CP violation is $\arg(\vev{S_3}^*\vev{S_1}^3
\vev{S_2})$, all these coefficients are {\it real}.

For the quarks,
\eqn\Msupd{M^d\sim\vev{\phi_d}\pmatrix{
\eb^3&0&\ea^3\eb+\ef\cr 0&\ea^4&\eb^2\cr 0&0&\eb\cr},}
\eqn\Msupu{M^u\sim\vev{\phi_u}\pmatrix{
\ea^2\eb^2&\ea^4&\ea^3\cr 0&\ea\eb&\eb\cr 0&\ea&1\cr}.}
There are no higher order corrections
to these entries because there is no holomorphic
combination of breaking parameters that is $\H$-invariant.

For the diagonal blocks of the squark mass-squared matrices,
\eqn\MLLd{\tilde M^{q2}_{LL}\sim\tilde m^2\pmatrix{
1&\ea^3\eb+\ef^*\eb^2&\ea^3+\ef^*\eb\cr
\ea^3\eb+\ef\eb^2&1&\eb\cr \ea^3+\ef\eb&\eb&1\cr},}
\eqn\MRRd{\tilde M^{d2}_{RR}\sim\tilde m^2\pmatrix{
1&\ea^7\eb^4&\ea^3\eb^2+\ef\eb^3\cr
\ea^7\eb^4&1&\ea^4\eb^2+\ef^*\ea\eb^3\cr
\ea^3\eb^2+\ef^*\eb^3&\ea^4\eb^2+\ef\ea\eb^3&1\cr},}
\eqn\MRRu{\tilde M^{u2}_{RR}\sim\tilde m^2\pmatrix{
1&\ea^2\eb^2&\ea\eb^2\cr \ea^2\eb^2&1&\ea\cr
\ea\eb^2&\ea&1\cr},}
where $\tilde m$ is the SUSY breaking scale. In each entry,
we explicitly wrote the subleading contributions to $\O(\lambda^4)$.

The mass-squared matrices for squarks that arise from
the $A$ terms are similar in form to the quark mass matrices.
Both $M^q$ and $\tilde M^{q2}_{LR}$ get non-holomorphic
contributions when the kinetic terms are rescaled to the canonical form
\Sequel. However, there are two additional (in general, non-holomorphic)
contributions to the {\it effective} $(\tilde M^{q2}_{LR})_{\rm eff}$
matrices: first, terms in the Kahler potential with one power of the
SUSY breaking spurion $\eta\equiv\tilde m\theta^2$ and, second,
insertions of the soft masses $\tilde M^{q2}_{LL,RR}$ on virtual squark
lines. All these sources can effectively be accounted for by estimating
$[(\tilde M^{q2}_{LR})_{\rm eff}]_{ij}\sim
[\tilde M^{q2T}_{LL}M^q\tilde M^{q2}_{RR}]_{ij}$:
\eqn\MLRd{(\tilde M^{d2}_{LR})_{\rm eff}\sim\tilde m\vev{\phi_d}\pmatrix{
\eb^3&\ea^7\eb+\ef\ea^4\eb^2&\ea^3\eb+\ef\cr
\ea^3\eb^4+\ef^*\eb^5&\ea^4&\eb^2\cr
\ea^3\eb^3+\ef^*\eb^4&\ea^4\eb&\eb\cr},}
\eqn\MLRu{(\tilde M^{u2}_{LR})_{\rm eff}\sim\tilde m\vev{\phi_u}\pmatrix{
\ea^2\eb^2+\ef\ea\eb^3& \ea^4+\ef\ea\eb& \ea^3+\ef\eb\cr
\ea\eb^3&\ea\eb&\eb\cr \ea\eb^2&\ea&1\cr}.}
In each term of \MLRd\ and \MLRu\ we wrote subleading contributions
(to $\O(\lambda^4)$) only to the extent that they carry non-trivial
phases (namely, are $\ec$-dependent).
Note that the zeros of \Msupd, \Msupu\ are lifted.
The non-holomorphic corrections are often ignored in the literature,
but they are important for CP violation.

The mass matrices given above lead \Sequel\ to the observed hierarchy
in quark masses and mixing angles;
highly suppress supersymmetric contributions
to $\Delta m_K$ and $\epsilon_K$; and induce $D-\bar D$
mixing which is close to the experimental bound.
In addition, as $(M^d)_{13}$ carries a phase of order 1,
the CKM matrix is complex with $\delta_{KM}=\O(1)$.

The gauginos do not transform under the horizontal symmetry.
Therefore, their masses are not suppressed by any of the
small parameters. Moreover, they are real, since
all holomorphic $\cal H$ invariants vanish.

The combination $\phi_u\phi_d$
carries $\H$-charge $(-1,0)$. Consequently, the $\mu$-term cannot arise
from the superpotential (it cannot be holomorphic in both
$\ea$ and $\ef$). It can still arise from the Kahler potential.
The selection rules imply then
\eqn\muandB{\mu\sim\tilde m\ea^*,\ \ \ m_{12}^2\sim\tilde m^2\ea^*.}
The magnitude of the $\mu$ term shows that the
Supersymmetric $\mu$ problem is solved. This is a specific
realization of the solution suggested in ref.
\ref\Nirmu{Y. Nir, Phys. Lett. B354 (1995) 107.}.
We note that this scenario implies $\tan\beta\sim5$, and
requires fine tuning of order $\lambda$ to get the
correct $m_Z/\tilde m$
\nref\RaNe{A.E. Nelson and L. Randall, Phys. Lett. B316 (1993) 516.}%
\nref\RaSa{R. Rattazzi and U. Sarid, Phys. Rev. D53 (1996) 1553.}%
\refs{\RaNe-\RaSa}.

\newsec{The Electric Dipole Moment of the Neutron}
We now consider the various contributions to $d_N$. 
The same analysis, with similar conclusions,
applies to the broader class of nuclear electric 
dipole moments.
(We use the calculations of ref.
\ref\FPT{W. Fischler, S. Paban and S. Thomas,
 Phys. Lett. B289 (1992) 373.}.)
First, we examine phases of the type $\phi_A$. In our
framework, where the $A$ terms are not proportional to
the Yukawa terms, there are many phases of this type.
The ones that are most crucial for $d_N$ are
\eqn\phiuA{
\phi_A^u=\arg\left({[V_L^u (M^u)_{\rm eff} V_R^{u\dagger}]_{11}\over
[V_L^u(\tilde M^{u2}_{LR})_{\rm eff} V_R^{u\dagger}]_{11}}\right),}
and the similarly defined $\phi_A^d$. In \phiuA,  $V_L^u$ and $V_R^u$
are the diagonalizing matrices for $(M^u)_{\rm eff}$
($V_L (M^u)_{\rm eff} V_R^{u\dagger}=M^u_{\rm diag}$);
$(M^u)_{\rm eff}$ is the up-quark mass matrix in the basis where the
kinetic terms are canonically normalized; and
$(\tilde M^{u2}_{LR})_{\rm eff}$ takes into account both the rotation
to this basis and the other contributions discussed in the previous
section. Indeed, the effect of the diagonalizing matrices
$V_M^q$ can be considered a fourth source of non-analytic
contributions to the effective $A$-terms. While the various entries
of $V_M^q$ can be thought of as carrying $\H$-charges similar to
$\tilde M^{q2}_{MM}$, one cannot apply quite the same selection rules,
since $V^q$ depends also on inverse powers of the $\epsilon_i$'s.

The experimental bound on $d_N$ requires 
 that $\phi_A^u,\phi_A^d\leq\O(10^{-2})$.
Examining eqs. \Msupd, \Msupu\ \MLRd\ and \MLRu, we find
\eqn\phiAud{\phi_A^u,\ \phi_A^d=\O(\lambda^6)\sim6\times10^{-5}.}
The reason for this strong suppression is that, to a good approximation,
the relevant Yukawa coupling and the corresponding $A$ term
are dominated by one and the same combination of
breaking parameters and, therefore, there is no
relative phase between them. (Remember that the $\O(1)$
coefficients in each of them are real.)
Therefore, the SUSY contribution to $d_N$ from this source is about
two orders of magnitude below the experimental bound.

Heavy quarks may contribute to $d_N$ through two-loop
diagrams. The relevant phases are $\phi_A^c$, etc.
These have to be smaller than $\sim10^{-1}$, and we
find that in our model they indeed are $\lsim\O(\lambda^6)$.
Their effects are therefore smaller than those of \phiAud.

Another possible source of large contribution to $d_N$
is a relative phase between the $\mu$ parameter of the
superpotential and the SUSY breaking $m_{12}^2$ term
in the scalar potential. Eq. \muandB\ reveals that there is no relative
phase between $\mu$ and $m_{12}^2$ and, therefore, no
contribution to $d_N$ from this source. More precisely,
both $\mu$ and $m_{12}^2$ get additional contributions
of order $\tilde m\ea^2\eb\ec^*$ and $\tilde m^2\ea^2\eb\ec^*$,
respectively. Consequently,
\eqn\phiB{\phi_B=\O(\ea\eb\ec)\sim2\times10^{-6}}
which is safely below the $d_N$ bound.

We conclude: in our model, combining an Abelian horizontal
symmetry and spontaneous CP violation, all the supersymmetric
contributions to FCNC processes and to CP violating quantities
are suppressed below the experimental bounds. $D-\bar D$
mixing is very close to the bound, while $d_N$ is about two
orders of magnitude below the bound.

\newsec{The Higgs Potential}
We would like to show that spontaneous CP violation
could naturally arise in our framework. For that purpose,
we have to introduce yet another Standard Model gauge singlet,
$S_4$, to which we assign $\H$-charge $(6,2)$.
The most general superpotential involving the $S_i$ fields is
\eqn\WofS{W(S_i)\sim aS_4S_3^2+{b\over\mpl^3}S_4S_3S_1^3S_2
+{c\over\mpl^6}S_4S_1^6S_2^2+\cdots,}
where the ellipses stand for terms with higher powers of
$S_4/\mpl$; $a,b,c$ are $ \O(1)$ coefficients.
Requiring $F_{S_1}=F_{S_2}=F_{S_3}=0$ can be solved
by $\vev{S_4}=0$. This prevents any change in our analysis
of quark and squark mass matrices of the previous section
due to the introduction of $S_4$. On the other hand, $F_{S_4}=0$
leads to
\eqn\minimum{a\ef^2+b\ea^3\eb\ef+c\ea^6\eb^2=0\ \Longrightarrow\
{\ef\over\ea^3\eb}={-b\pm\sqrt{b^2-4ac}\over2a}.}
We learn that
\item{(i)} The ratio of VEVs that we used, $\ef\sim\ea^3\eb$,
arises naturally as a consequence of the $\H$-charge assignments;
\item{(ii)} A relative complex phase between $\vev{S_3}$
and $\vev{S_1}^3\vev{S_2}$ arises for $b^2-4ac<0$. This is similar to
the mechanism used in \BaBb.

The overall scale of the VEVs $\vev{S_i}$ is not determined
by \minimum. However, it is attractive to assume that the
horizontal $U(1)$'s are gauged and have anomalies
\ref\IbRo{L. Ibanez and G.G. Ross, Phys. Lett. B322 (1994) 100.}\
which are cancelled
by the Green-Schwarz mechanism
\ref\GrSc{M. Green and J. Schwarz, Phys. Lett. B149 (1984) 117.}.
This indeed requires that the scales $\vev{S_i}$ are not
far below $\mpl$
\nref\DSW{M. Dine, N. Seiberg and E. Witten,
 Nucl. Phys. B289 (1987) 589.}%
\nref\ADS{J. Atick, L. Dixon and A. Sen, Nucl. Phys. B292 (1987) 109.}%
\nref\DIS{M. Dine, I. Ichinose and N. Seiberg,
 Nucl. Phys. B293 (1987) 253.}%
\refs{\DSW-\DIS}.

\newsec{Conclusions}
The combination of Abelian horizontal symmetries and
spontaneous CP violation could give viable models where,
without any fine tuning, the following features arise:
\item{(a)} Quark masses and mixing angles exhibit the
observed smallness and hierarchy;
\item{(b)} Supersymmetric contributions to FCNC are
suppressed. The suppression of the contribution to
$K-\bar K$ mixing is satisfactory only in a special
class of models, where the Cabibbo angle is generated
in the up sector and not in the down sector;
\item{(c)} The $\mu$ problem is solved: the $\mu$ term
arises only from the Kahler potential and is somewhat
below the supersymmetry breaking scale;
\item{(d)} Supersymmetric contributions to CP violation
and, in particular, to the electric dipole moment of the
neutron, are suppressed below the experimental bounds.

Our solution is particularly relevant for squark masses of a
few hundred GeVs. If squarks are very light, {\it i.e.} $\sim100\
\gev$, then
the quark-squark alignment solution of the $\Delta m_K$
problem runs into problems with the $\Delta m_D$ bound. (Notice
however that the gluino dominance of the RG evolution 
at low energy can easily
induce  a mild  $\O(10\%)$ degeneracy among the
squarks of the first two families.
This allows to naturally satisfy the bounds on $D-\bar D$ mixing
for squark masses as low as $\sim 200$ GeV.)
If, on the other hand, squark masses are $\sim1\ \tev$,
then the suppression of SUSY contributions to $d_N$ relaxes
the requirement that CP violating phases are small.

We have not addressed the strong CP problem in this work.
We would like to mention, however, that horizontal
symmetries can {\it naturally} solve this problem by setting the
bare $m_u$ to zero
\ref\BNS{T. Banks, Y. Nir and N. Seiberg, in {\it Yukawa couplings
and the origin of mass}, ed. P. Ramond (International Press,
Cambridge MA, 1995), hep-ph/9403203.}. Alternatively, there
does not seem to be any theoretical obstacle in this
scenario to having an axion solution. 

Let us now comment on the generality of our mechanism
for solving the Supersymmetric CP problem.
The suppression of the Supersymmetric contribution to $\epsilon_K$
is satisfactory in all models of quark-squark alignment
\Sequel. However, this is not the case for $d_N$.
Here, additional conditions, all realized in our specific
model, should hold:
\item{(i)} Some entries in the Yukawa matrices should
get comparable contributions from two combinations of
breaking parameters which differ in their phases.
\item{(ii)} The Yukawa coupling $Y^d_{11}$ and $Y^u_{11}$
should each be dominated by a single combination of small parameters.
\item{(iii)} The contribution to the Yukawa couplings
of the up and down quark mass eigenstates should be
dominated (to high order in the small breaking parameter)
by the respective $Y_{11}$ couplings;
alternatively, other contributions (e.g. $Y_{12}Y_{21}/Y_{22}$)
should carry the same phase as $Y_{11}$.
\item{(iv)} The $\mu$ term should
be dominated by a single combination of small parameters.
\item{(v)} The non-holomorphic contributions to the effective Yukawa-
and $A$-terms should either carry no non-trivial phases or be very small.

Condition (i) is necessary in order to get a non-zero
CKM phase. In our framework of quark-squark alignment,
the Standard Model box diagram is the only possible source
of $\epsilon_K$ and, therefore, $\delta_{KM}=\O(1)$ is necessary.
Condition (ii) is necessary in order that
$\phi_A^{u,d}$ is not of order 1. To implement (i) in our
framework, the charges of $S_3$ have to be such that it contributes
at leading order to at least one of the entries in
$M^d$ or $M^u$. On the other hand, (ii) requires
that it does not contribute at leading order to $Y_{11}^q$.
This leaves only a limited set of possible charges for $\H(S_3)$.
With our assignments of $\H$ charges for quarks, these are
$(-3,0)$, $(-4,0)$ and $(-3,-1)$.
Condition (iii) is achieved when holomorphy requires $Y_{21}=Y_{31}=0$
in both sectors (which is easy to implement) but also puts further
restrictions on $\H(S_3)$, because $Y^u_{12}$ should be real to a good
approximation. If  $S_3$  does contribute to $Y^u_{12}$ (as would be the
case with $S_3(-3,0)$ or $S_3(-4,0$)), the phase $\phi_A^u$ is still
suppressed, but only by $\O(\lambda^2)$. We employ $S_3(-3,-1)$,
which gives the strongest possible suppression of $\phi_A^u$ (while
keeping the CKM matrix complex). Condition (iv) is necessary to
assure that $\phi_B$ is not  $\O(1)$.
Condition (v) imposes various requirements, in particular
that the leading $\H$-invariant combination of breaking parameters
(in our case, $\ea^3\eb\ec^*$) is extremely small. 
However  (v) requires in general more than just that. 
For instance, when $M^d_{13}$ 
carries a non trivial phase, it constrains
 $(\tilde M^{d2}_{RR})_{31}$ to be relatively small, limiting the 
choices of $\cal H$ quantum numbers for the quarks. 
Both (iv) and (v) are usually satisfied, at least at
$\O(\lambda^2)$, once the other conditions are fulfilled.

To summarize:
\item{1.} Models of quark squark alignment have a satisfactory
suppression of supersymmetric contributions to $\epsilon_K$.
\item{2.} When combined with spontaneous CP violation, these
models give, in general, a satisfactory suppression of $\phi_B$.
\item{3.} In a class of these models, with specific choices of
the horizontal charges of the scalar whose VEV breaks CP,
the $\phi_A$ phases are suppressed but $\delta_{KM}$ is not.
\item{4.} With an almost unique choice of these charges,
the suppression of $\phi_A$ is well below the experimental bound.
Otherwise, $\phi_A$ is close to the bound (but may be acceptable,
taking into account the large theoretical uncertainties in the
calculation of $d_N$).

More generally, we believe that the basic idea, namely that
CP violating phases arise only in connection with small
breaking parameters of a horizontal symmetry, might be
useful in solving or relaxing the SUSY CP problem.

\vskip 1cm
{\bf Acknowledgements:}
We would like to thank K.S. Babu, M. Dine and R. Leigh for
useful discussions.
YN is grateful to the New High Energy Theory Center of Rutgers University
and to the School of Natural Sciences of the Institute for Advanced
Studies in Princeton for their hospitality.
YN is supported in part by the United States -- Israel Binational
Science Foundation (BSF), by the Israel Commission for Basic Research,
and by the Minerva Foundation (Munich).
The work of RR is supported by the National Science Foundation
under grant PHY-91-21039.

\listrefs
\end